\documentclass{aa}  

\usepackage{natbib}
\usepackage{graphicx}
\usepackage{amsmath}
\newcommand*{\dittoclosing}{---''---}

\usepackage{color}

\begin{document} 

\bibliographystyle{aa}

\title{Compositional characterisation of the Themis family}

   \author{M. Marsset\inst{1,2}
          \and
          P. Vernazza\inst{2}
          \and
          M. Birlan\inst{3,4}
          \and
          F. DeMeo\inst{5}
          \and
          R. P. Binzel\inst{5}
          \and
          C. Dumas\inst{1}
          \and
          J. Milli\inst{1}
          \and
          M. Popescu\inst{4,3}
          }

   \institute{European Southern Observatory (ESO),
              Alonso de C\'ordova 3107, 1900 Casilla Vitacura, Santiago, Chile\\
              \email{mmarsset@eso.org}
        \and
         Aix Marseille University, CNRS, LAM (Laboratoire d'Astrophysique de Marseille) UMR 7326, 13388, Marseille, France
        \and
        IMCCE, Observatoire de Paris, 77 avenue Denfert-Rochereau, 75014 Paris Cedex, France
        \and
        Astronomical Institute of the Romanian Academy, 5 Cu\c{t}itul de Argint, 040557 Bucharest, Romania
        \and
        Department of Earth, Atmospheric and Planetary Sciences, MIT, 77 Massachusetts Avenue, Cambridge, MA, 02139, USA \\
        }

   \date{Received; accepted}

 
  \abstract
   {It has recently been proposed that the surface composition of icy main-belt asteroids (B-, C-, Cb-, Cg-, P-, and D-types) may be consistent with that of Chondritic porous interplanetary dust particles (CP IDPs).}
   {In the light of this new association, we re-examine the surface composition of a sample of asteroids belonging to the Themis family in order to place new constraints on the formation and evolution of its parent body.} 
   {We acquired near-infrared spectral data for 15 members of the Themis family and complemented this dataset with existing spectra in the visible and mid-infrared ranges to perform a thorough analysis of the composition of the family. Assuming end-member minerals and particle sizes ($<2\,\mu$m) similar to those found in CP IDPs, we used a radiative transfer code adapted for light scattering by small particles to model the spectral properties of these asteroids.} 
   {Our best-matching models indicate that most objects in our sample (12/15) possess a surface composition that is consistent with the composition of CP IDPs. We find ultra-fine grained ($<2\,\mu$m) Fe-bearing olivine glasses to be among the dominant constituents. We further detect the presence of minor fractions of Mg-rich crystalline silicates (enstatite and forsterite). The few unsuccessfully matched asteroids may indicate the presence of interlopers in the family or objects sampling a distinct compositional layer of the parent body.}
   {The composition inferred for the Themis family members suggests that the parent body accreted from a mixture of ice and anhydrous silicates (mainly amorphous) and subsequently underwent limited heating. By comparison with existing thermal models that assume a 400-km diameter progenitor, the accretion process of the Themis parent body must have occurred relatively late ($>$4 Myrs after CAIs) so that only moderate internal heating occurred in its interior, preventing aqueous alteration of the outer shell.}
   

   \maketitle
%

\section{Introduction}
\label{sec:introduction}

Discovered in 1918 \citep{Hirayama:1918jb} and located in the outer belt around $\sim$3.12 AU,  the Themis family comprises more than 4~000 dynamically well-established members \citep{Nesvorny:2012abc}. The family is predicted to have formed $\sim$2.5 $\pm$ 1.0 \,Gyrs ago \citep{Broz:2013jl} following the catastrophic collisional disruption of a $\sim$270\,km \citep{Broz:2013jl} or a $\sim$400\,km-sized body \citep{Marzari:1995ga, Durda:2007im}. Observations over the past ten years have revealed that this family is relatively unique for a number of reasons: \\

- Observations in the visible and mid-infrared wavelength ranges have shown that it mainly consists of low-albedo (${\rm p_v<0.1}$; \citealt{Masiero:2011jc, Masiero:2014gu}) B- and C-type asteroids \citep{Florczak:1999et, MotheDiniz:2005hma, Clark:2010jo, Ziffer:2011iga, deLeon:2012ega, Fornasier:2014uu, Kaluna:2016jo}. In parallel, similar observations combined with numerical simulations have shown that there is no other B- or C-type family derived from the catastrophic disruption of a large (D>200~km) body in the asteroid belt \citep{Durda:2007im, Broz:2013jl}. As such, the Themis family offers us a unique opportunity to investigate the internal compositional structure of a large B/C-type asteroid, which in turn can allow us to put constraints on the early thermal evolution of such asteroid types. 

- Several main-belt comets have been dynamically linked to the Themis family \citep{Hsieh:2006dk, Hsieh:2009gy, Novakovic:2012bf}.

- (24)~Themis -- the largest member of the family -- is the first main belt asteroid for which water ice has been unambiguously detected at the surface \citep{Campins:2010fi, Rivkin:2010ge}. Since then, similar water ice detections have been reported for other asteroids, including four additional Themis family members \citep{Takir:2012cza, Hargrove:2015ii}.
 
- The density values recorded for two family members (90~Antiope and 379~Huenna: <1.3\,g/cm$^3$; \citealt{Descamps:2007im, Marchis:2008eq}) are among the lowest densities measured so far for large (D>100\,km) main-belt asteroids, very likely implying a high fraction of ice(s) in the interior of these bodies (see discussion on the porosity and ice/rock composition of 90~Antiope in \citealt{CastilloRogez:2010kya}).

The last three features all point towards an ice-rich composition for the Themis parent body while showing little compatibility with a thermally metamorphosed parent body that has been heated throughout at temperatures exceeding 300\,K. The latter interpretation has been previously suggested on the basis of a similarity between the near-infrared spectral properties of heated CI/CM chondrites and those of the family members' surfaces \citep{Clark:2010jo, Ziffer:2011iga}. In brief, and as in the case of other B- and C-type asteroids \citep{Vernazza:2015ei}, the Themis family members appear unsampled by our meteorite collections. Interplanetary dust particles (IDPs) may instead be more appropriate extraterrestrial analogues for these objects' surfaces \citep{Vernazza:2015ei}. 

IDPs are heterogeneous aggregates of predominantly sub- to micrometre-sized crystalline silicate grains (e.g., olivine, pyroxenes), silicate glasses, iron-rich sulfides (e.g., pyrrhotite, troilite), and FeNi alloys incorporated into a carbonaceous matrix (predominantly amorphous and partially graphitic) \citep{Messenger:2003cz}. IDPs are usually classified into three broad classes: the olivine-rich IDPs, the pyroxene-rich IDPs, and the layer lattice silicate-rich (hereafter "hydrous") IDPs \citep{Sandford:1989gr}. The first two classes, composing the family of chondritic porous (CP) IDPs, have recently been associated with P-, D-type asteroids and comets (olivine-rich IDPs), and B-, C-, Cb-, and Cg-type asteroids (hereafter "BCG"; pyroxene-rich IDPs) \citep{Vernazza:2015ei}. 

In light of these recent associations, we investigate in this paper the surface mineralogy of a sample of Themis family members using a combined dataset of visible, near-infrared, and mid-infrared spectra. Specifically, we acquired near-infrared spectral data for 15 members of the Themis family (section~\ref{sec:observations}), thus expanding the number of surveyed objects in this wavelength range by a factor of 2 \citep{Ziffer:2011iga, deLeon:2012ega}. We complemented this dataset with existing spectra in the visible \citep{Bus:2002ia, Lazzaro:2004ja} and mid-infrared \citep{Licandro:2012kt, Hargrove:2015ii} ranges to perform a thorough analysis of the composition of the Themis family. Multi-wavelength spectral analysis has proven to be a powerful tool for constraining the surface composition of asteroids. For many asteroid classes that are relatively featureless in the visible and near-infrared (VNIR) wavelength ranges (e.g., the P- and D-classes), mid-infrared (MIR) spectroscopy provides a nice complement. Indeed, most major mineral groups and silicate glasses that lack useful diagnostic features in the VNIR range do exhibit diagnostic MIR features (e.g., \citealt{Emery:2006ij, Vernazza:2012kv}). Assuming end-member minerals and particle sizes (typically sub- to micrometre sizes) similar to those found in CP IDPs, we use a radiative transfer code adapted for light scattering by small particles compared to the wavelength to model the spectral properties of the Themis family members (section~\ref{sec:composition}). Finally, we place the results of our compositional analysis in the context of the formation and thermal evolution of the Themis parent body (section~\ref{sec:discussion}). 


\section{Observation and data reduction}
\label{sec:observations}
The new data presented here are near-infrared spectra for 15 Themis family members collected between February 2010 and January 2012 using the 3-m NASA IRTF (Mauna Kea, HI; see Table~\ref{table:phys}). Observing runs were conducted remotely primarily from the Observatory of Paris-Meudon (France) \citep{Birlan:2004gu}. The low- to medium-resolution spectrograph and imager SpeX \citep{Rayner:2003hfa}, combined with a 0.8$\times$15" slit, was used in the low-resolution prism mode to acquire the spectra in the 0.7-2.5~$\mu$m wavelength range. To monitor the high luminosity and variability of the sky in the near-infrared, the telescope was moved along the slit during the acquisition of the data so as to obtain a sequence of spectra located at two different positions (A and B) on the array. These paired observations provided near simultaneous sky and detector bias measurements. Objects and standard stars were observed near the meridian to minimise their differences in airmass and match their parallactic angle to the fixed N/S alignment of the slit. Our primary solar analogue standard stars were Hyades~64 and Land~98-978. Additional solar analog stars with comparable spectral characteristics were used around the sky. 

Reduction was performed using a combination of routines within the Image Reduction and Analysis Facility (IRAF), provided by the National Optical Astronomy Observatories (NOAO) \citep{Tody:1993vm}, and the Interactive Data Language (IDL). We used a software tool called "autospex" to streamline reduction procedures, which included bad-pixel mapping and correction, flat-fielding, subtraction of the AB dithered pairs, extraction of the 2D spectra, and wavelength calibration. Using IDL, an absorption coefficient based on the atmospheric transmission (ATRAN) model by \citet{Lord:1992to}, was determined for each object and star pair that best minimises atmospheric water absorption effects, in particular near 1.4 and 2.0\,$\mu$m, the locations of major absorption bands due to telluric H$_2$O. The final IDL step averaged all the object and standard pairs to create the final reflectance spectrum for each object.

\begin{table*}[tbh]
\small
\begin{center}
\caption[]{\em{\small Physical and spectral properties of the targeted asteroids\\}}
\label{table:phys}
\begin{tabular}{llcccccc}
\hline \hline \noalign {\smallskip}
No. & Name & Observing date & Tax. & p$_{\rm v}$ & {\it D} & Spectral slope & Perihelion \\
 & & & \citep{DeMeo:2009gza} & & (km) & (\% $\mu$m$^{-1}$) & (AU) \\
\hline \noalign {\smallskip}
24 & Themis & 2005-10-08 & B & 0.064$\pm$0.016 & 202$\pm$6 & 8.1$\pm$0.3 & 2.74 \\
62 & Erato & 2009-09-20 & B & 0.061$\pm$0.003* & 95$\pm$2* & 2.8$\pm$0.3 & 2.59 \\
90 & Antiope & 2005-09-05 & C & 0.057$\pm$0.012 & 121$\pm$2 & 20.0$\pm$0.3 & 2.63\\
171 & Ophelia & 2010-02-22 & C & 0.077$\pm$0.020 & 104$\pm$1 & 13.2$\pm$0.2 & 2.72 \\
223 & Rosa & 2011-10-19 & T & 0.034$\pm$0.005 & 83$\pm$3 & 42.4$\pm$0.2 & 2.74 \\
268 & Adorea & 2009-09-20 & Cb & 0.044$\pm$0.007 & 141$\pm$3 & 28.7$\pm$0.2 & 2.67 \\
316 & Goberta & 2010-09-07 & C & 0.059$\pm$0.007 & 60$\pm$1 & 9.2$\pm$0.1 & 2.74 \\
379 & Huenna & 2010-02-22 & C & 0.065$\pm$0.008 & 87$\pm$2 & 9.8$\pm$0.3 & 2.55 \\
383 & Janina & 2010-02-22 & B & 0.096$\pm$0.013 & 45$\pm$1 & -4.2$\pm$0.3 & 2.63 \\
461 & Saskia & 2012-01-25 & X & 0.048$\pm$0.008 & 49$\pm$0 & 24.3$\pm$0.1 & 2.68 \\
468 & Lina & 2012-01-21 & X & 0.050$\pm$0.009 & 65$\pm$2 & 28.5$\pm$0.2 & 2.51 \\
526 & Jena & 2011-10-25 & C & 0.058$\pm$0.018 & 51$\pm$1 & 2.9$\pm$0.1 & 2.70 \\
621 & Werdanti & 2010-11-01 & B & 0.152$\pm$0.018* & 27$\pm$2* & 4.2$\pm$0.2 & 2.69 \\
767 & Bondia & 2010-05-11 & B & 0.096$\pm$0.018 & 43$\pm$1 & -2.4$\pm$0.3 & 2.58 \\
954 & Li & 2010-09-04 & Cb & 0.055$\pm$0.007 & 52$\pm$1 & 17.4$\pm$0.1 & 2.59 \\
\hline \noalign {\smallskip}
\end{tabular}
\begin{flushleft}
{\bf Notes.}  ${\rm p}_{\rm v}$: optical albedo; {\it D}: effective diameter. Albedo and diameter values are from the WISE/NEOWISE database \citep{Masiero:2011jc}. Values for objects that were not observed by WISE are indicated by an asterisk and are from the Planetary Data System (NASA)\\
\end{flushleft}
\end{center} 
\end{table*}


\section{Spectral mixing model}
\label{sec:composition}

The spectral analysis of the Themis members is made in two steps. First (section~\ref{sec:mir}), we analyse the silicate composition of a subset of the objects using available MIR data from the literature. In the case of surfaces covered by small grains, which is the hypothesis we investigate here, the MIR spectral range allows a more confident distinction between the components of a silicate-rich mixture than does the VNIR range. This is the reason for starting our analysis in this wavelength range. As a next step (section~\ref{sec:nir}), we use the silicate composition derived from the MIR as a bottom line for our spectral analysis in the VNIR. In the latter range, we use additional end members that are found in CP IDPs, such as sulfides, FeNi metals, and carbonaceous compounds. 

\begin{table*}[tbh]
\small
\begin{center}
\caption[]{\em{\small Standards used in the modelling procedure\\}}
\label{table:template}
\begin{tabular}{lccccc}
\hline \hline \noalign {\smallskip}
Component & Category & Chemical formula & Wavelength range & Type of data$^{*}$ & Reference \\
 & & & (~$\mu$m) & & \\
\hline \noalign {\smallskip}
Amorphous olivine & Silicate & Mg$_{2x}$Fe$_{2-2x}$SiO$_4$ & 0.3-2.5, 8.0-13.0 & OC & \citet{Dorschner:1995wq} \\
Amorphous pyroxene & \dittoclosing & Mg$_{x}$Fe$_{1-x}$SiO$_3$ & 0.3-2.5, 8.0-13.0 & OC & \citet{Dorschner:1995wq} \\
Crystalline forsterite & \dittoclosing & Mg$_2$SiO$_4$ & 0.3-2.5 & OC & \citet{Lucey:1998ko} \\
\dittoclosing & \dittoclosing & \dittoclosing & 8.0-13.0 & MAC & \citet{Koike:2006bo} \\
Crystalline fayalite & \dittoclosing & Fe$_2$SiO$_4$ & 8.0-13.0 & MAC & \citet{Koike:2006bo} \\
Crystalline enstatite & \dittoclosing & MgSiO$_{3}$ & 0.3-2.5 & OC & \citet{Lucey:1998ko} \\
\dittoclosing & \dittoclosing & \dittoclosing & 8.0-13.0 & OC & \citet{Zeidler:2015fu} \\
Crystalline diopside & \dittoclosing & CaMgSi$_2$O$_6$ & 8.0-13.0 & MAC & \citet{Koike:2000ub} \\
Pyrrhotite & sulfide & Fe$_{1-x}$S (0$<$x$<$0.2) & 0.3-2.5 & RS & USGS$^{**}$ \\
Troilite & \dittoclosing & FeS & 0.3-2.5 & OC & \citet{Semenov:2003hk} \\
Iron & Metal & Fe & 0.3-2.5 & OC & \cite{Semenov:2003hk} \\
Amorphous carbon & Carbon & C & 0.3-2.5 & OC & \citet{Zubko:1996wb} \\
\hline \noalign {\smallskip}
\end{tabular}
\begin{flushleft}
$^*$OC=optical constants ($n$ and $k$); MAC=mass absorption coefficient ($\kappa$); RS=reflectance spectrum (when OC not available). $^{**}$http://speclab.cr.usgs.gov\\
\end{flushleft}
\end{center} 
\end{table*}

\subsection{Mid-infrared range (8-13 microns)}
\label{sec:mir}
\citet{Licandro:2012kt} and \citet{Hargrove:2015ii} acquired spectra over the MIR spectral range for several Themis asteroids using the Spitzer Space Telescope. Silicate emissions were unambiguously detected in the cases of (24)~Themis, (383)~Janina, (468)~Lina, (515)~Athalia, and (526)~Jena. 

We modelled the objects' spectra through radiative transfer to constrain the composition of their silicates. The continuum-removed emissivity was fitted by the absorption efficiency factor ($Q_A$) of a distribution of small hollow spheres with equivalent radii 1~$\mu$m and a volume fraction occupied by the central vacuum inclusion ranging from 0\%\ to 90\% \citep{Min:2003km}. We used the optical constants (n and k) of common end-member silicates found in anhydrous chondritic porous IDPs (Bradley 2005) retrieved mainly from the Jena database\footnote{http://www.astro.uni-jena.de/Laboratory/Database/databases.html}. For crystalline silicates we used the Fe and Mg end members of olivine (forsterite: Mg$_2$SiO$_4$ and fayalite: Fe$_2$SiO$_4$) and the Mg end member and Ca--Mg solid solution of pyroxene (enstatite: MgSiO$_3$ and diopside: CaMgSi$_2$O$_6$). For amorphous silicates, we used the Fe and Mg end members of olivine and the Mg end member of pyroxene. Each fit to the asteroid spectrum was modelled as a linear combination of the end-member emissivity spectra whose relative abundances were the free parameters of our model. 

A limitation of the linear mixture approach is that it is strictly valid only when the end members are arranged in discrete patches at the surface, which is certainly not the case for asteroid surfaces. Nonetheless, we adopted this approach because it provides a qualitative estimate of the possible end members. Understanding how these end members are truly distributed and constraining their relative abundance precisely is beyond the scope of this paper.

Finally, an IDL routine was used to find the minimum root-mean-square (rms) residual between the measured spectrum and the computed spectrum.

When applying our fitting procedure to the objects' spectra, it first appears that amorphous grains are more abundant (>70\%) than crystalline ones (<30\%). In all cases but one (468 Lina; see discussion at section~\ref{sec:discussion}), including crystalline silicates helps to improve the quality of the fit. Second, it appears that crystalline silicates are in the Mg-rich form (enstatite and forsterite) and that the crystalline enstatite grains are more abundant than the crystalline forsterite ones (ol/(ol+py)$<$40\%), which agrees with the association between BCG-type asteroids and pyroxene-rich IDPs proposed by \citet{Vernazza:2015ei}. 

\begin{figure}[tbh]
\centering
\includegraphics[angle=0, width=\linewidth, trim=0cm 0.9cm 0.9cm 0cm, clip]{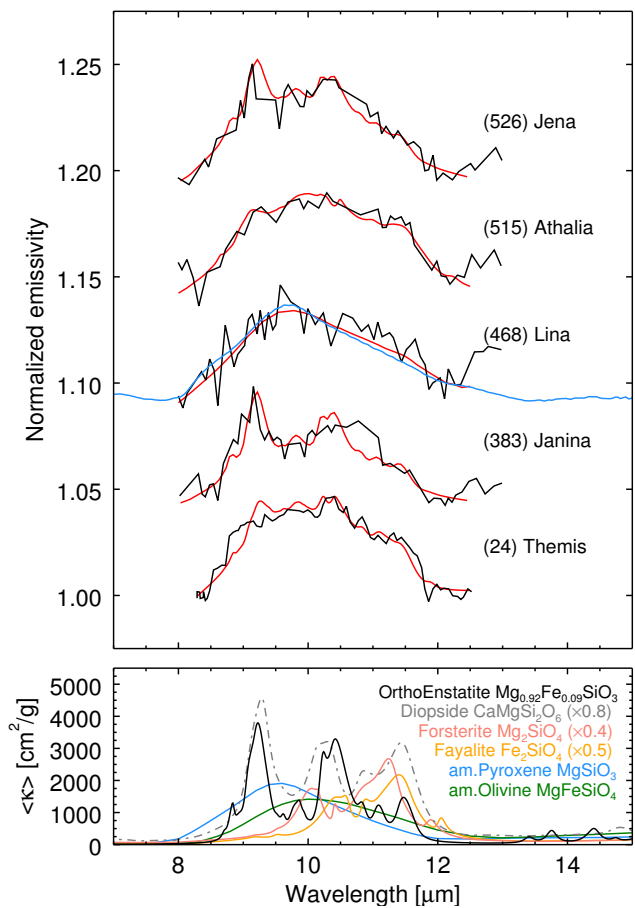}
 \caption{{\it Top:} Best fit (red) to the spectra of the Themis family members (black)  using a Rayleigh scattering model. The spectrum of asteroid (468)~Lina is well-matched by the spectrum of the hydrous IDP W7068A28MIR (blue; \citealt{Merouane:2014ge}; see discussion in section~\ref{sec:discussion}). {\it Bottom:} Mass absorption coefficients of the end members used in the fitting procedure. Diopside (dotted line), forsterite, and fayalite were scaled down to 80\%, 40\%, and 50\%, respectively.}
\label{fig:tir_spectra}
\end{figure}

The presence of crystalline enstatite is supported by the presence of two prominent emissivity features at $\sim$9.2~$\mu$m and $\sim$10.4~$\mu$m in the cases of (383)~Janina and (526)~Jena, and possibly at $\sim$10.4~$\mu$m in the case of (24)~Themis. Finally, the emissivity around 11.2~$\mu$m, where amorphous compounds become minor contributors to the total emissivity, is best reproduced when adding a small fraction of crystalline forsterite to the model. The best-fit solution for each asteroid and the relative contributions of the different dust species are listed in Table~\ref{table:compo_mir} and the corresponding model spectra are shown in Fig.~\ref{fig:tir_spectra}.

\begin{table*}[tbh]
\small
\begin{center}
\caption[]{\em{\small Silicate composition of each asteroid derived from the MIR spectra\\}}
\label{table:compo_mir}
\begin{tabular}{lcccccccc}
\hline \hline \noalign {\smallskip}
Asteroid & ol/(ol+px) & Mg/(Mg+Fe) & Ca/(Ca+Mg) & Amorphous content \\
\hline \noalign {\smallskip}
(24) Themis & 0.2 & 1.0 & 0.0 & 0.8 \\
(383) Janina & 0.1 & 1.0 & 0.0 & 0.7 \\
(468) Lina$^*$ & -- & -- & -- & 1.0 \\
(515) Athalia & 0.4 & 1.0 & 0.0 & 0.9 \\
(526) Jena & 0.1 & 1.0 &  0.0 & 0.8 \\
\hline \noalign {\smallskip}
\end{tabular}
\begin{flushleft}
{\bf Notes.} The olivine/(olivine+pyroxene) ratio is given for crystalline silicates, the Mg/(Mg+Fe) ratio for crystalline olivine, and the Ca/(Ca+Mg) ratio for crystalline pyroxenes, when present in the model. ${}^*$See discussion for this asteroid in section~\ref{sec:discussion}\\
\end{flushleft}
\end{center} 
\end{table*}

\subsection{Visible and near-infrared range (0.4-2.5 microns)}
\label{sec:nir}
\begin{figure}[tbh]
\centering
\includegraphics[angle=0, width=\linewidth, trim=0cm 0.5cm 0cm 0cm, clip]{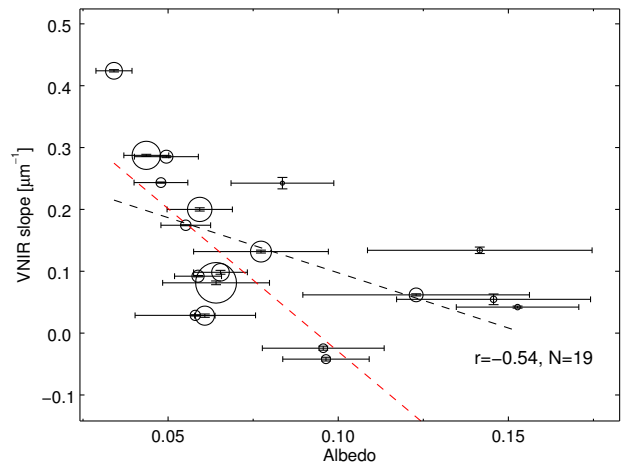}
\caption{VNIR spectral slopes versus albedos for 19 Themis family members. The size of the circles is proportional to the diameter of the objects. The black and red dotted lines correspond to a linear regression of the data points, unweighted (black line) and weighted by the uncertainty on the measurements (red line). A strong anti-correlation suggests that a single mechanism is responsible for bluer spectra and higher albedos. 
Data from \citet{Ziffer:2011iga} and \citet{deLeon:2012ega} and the present work were used to produce this figure.}
\label{fig:slope}
\end{figure}

We first performed a basic analysis of the reflectance spectra obtained with SpeX for 15 Themis family members and complemented in the visible by data from the SMASS \citep{Bus:2002ia} and S3OS2 \citep{Lazzaro:2004ja} surveys. 
This new dataset reveals a continuous trend in spectral slopes from blue ($\sim$-0.1\,$\mu$m$^{-1}$) to very red ($>$0.4\,$\mu$m$^{-1}$) ones. In addition, we find that the spectral slopes and albedos are anti-correlated, a trend that is reinforced when including additional spectra from \citet{Ziffer:2011iga} and \citet{deLeon:2012ega} (correlation factor $r=-0.54$ for a sample of $N=19$ objects, i.e., a confidence level $>$99.1\% that the correlation is not random; Fig.~\ref{fig:slope}). The correlation remains significant ($r=-0.36$ for $N=16$, i.e., a confidence level $>$91.5\% that the correlation is not random) when rejecting possible interlopers from our sample, namely the objects that possess a distinct spectral shape with respect to the other members (223~Rosa, 461~Saskia, and 468~Lina). Possible origins for this apparent compositional trend are discussed in section~\ref{sec:discussion}. 

As a next step, we modelled the spectral properties of the 15 Themis family members in order to i) verify the compositional solution derived from the MIR range and ii) possibly detect the presence of other compounds in addition to silicates (e.g., iron, carbon, sulfides). Modelling of the light reflected from particulate surfaces generally requires the use of geometric-optics equations in the context of Hapke's scattering theory \citep{Hapke:1993tt}. This approach, however, is only valid when the size of the particles is large compared to the wavelength, which is not the case for the typical IDPs' building blocks. 
Previous works \citep{Moersch:1995ip, Mustard:1997ey, Pitman:2005hk} have shown that a combination of the Mie \citep{Bohren:1983wi} and Hapke \citep{Hapke:1993tt} theories provides a more appropriate means of calculating the reflectance and emittance of a surface covered by small grains than geometric optics alone. We therefore adopted a similar approach in our calculation of the asteroid reflectances.

A detailed description of our code can be found in Appendix~\ref{sec:app}. In brief, we used the optical constants available in the literature for eight species. These include the silicate end members found to contribute to the 10~$\mu$m-band emission (amorphous olivine and pyroxene and crystalline enstatite and forsterite ), as well as the main non-silicate end members found in CP IDPs, namely sulfides (pyrrhotite: Fe$_{1-x}$S (0$<$x$<$0.2), troilite: FeS), iron, and amorphous carbon (Table~\ref{table:template}). We computed the single-scattering albedo ($w_\lambda$) for each specie using geometric-optics equations for particles much larger than the wavelength (arbitrary defined as $d>4.5\lambda$, where d is the equivalent diameter of the grain), and Mie calculations for particles smaller than or comparable in size to the wavelength ($d<1.5\lambda$). 
Since the transition between the Mie theory and geometric optics is rather arbitrary, we defined a transition regime where $w_\lambda$ is calculated as a weighted average of the two theories. 

\begin{figure*}[tbh]
\sidecaption
\centering
\includegraphics[angle=0, width=12cm, trim=0.6cm 0.5cm 2cm 0cm, clip]{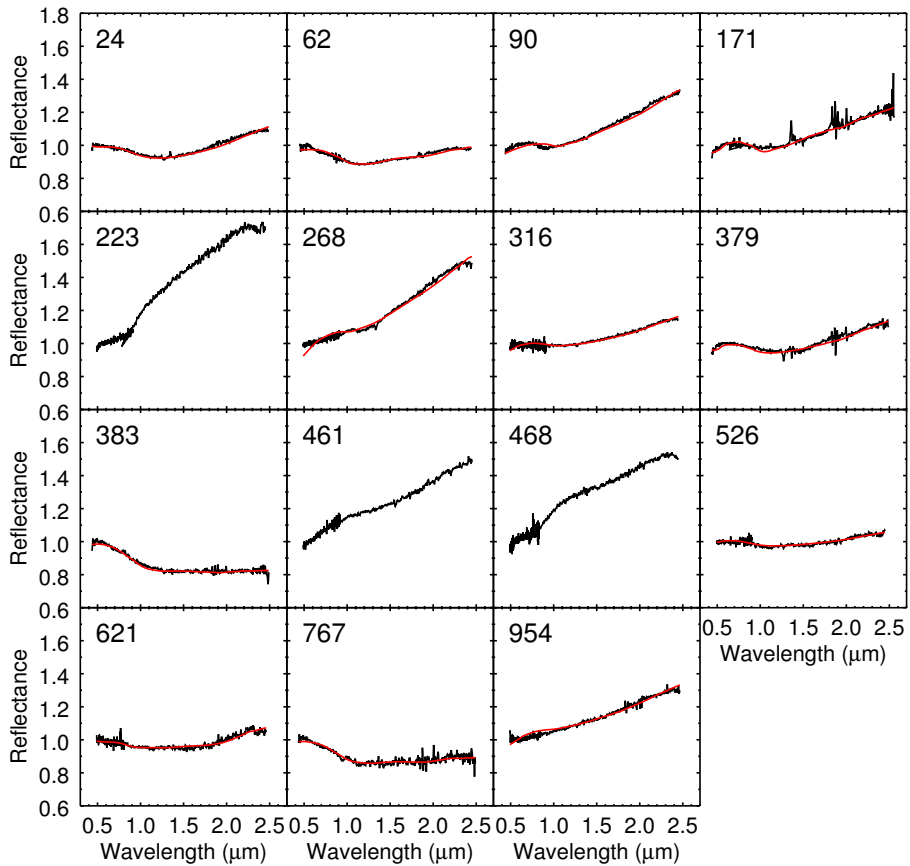}
 \caption{Comparison between the VNIR spectra of the Themis family members (black) and the best-fit models (red) obtained using the Mie-Hapke scattering model. Asteroid spectra are normalised to unity at 0.55~$\mu$m.}
\label{fig:nir_spectra}
\end{figure*}

Finally, the reflectance was derived from $w_\lambda$ for each grain size and averaged over the full size distribution of the particles. 
Typically, we assumed a power-law distribution (${\rm d}n(\alpha)=\alpha^p{\rm d}\alpha, p<0$) that is consistent with a collisionally evolved regolith resulting from micrometeoroid bombardment and covering the size range of the IDPs' building blocks (0.1-2~$\mu$m). 
Each fit to the asteroid spectrum was modelled as a linear combination of the end-member reflectance spectra whose relative abundances were the free parameters of our model. Finally, an IDL routine was used to find the minimum rms residual between the measured spectrum and the computed spectrum. Overall, good-matching fits were found for all the asteroids in our sample, except for (223)~Rosa, (461)~Saskia, and (468)~Lina (see discussion in section~\ref{sec:discussion}). The best-fit solution for each asteroid is provided in Table~\ref{table:compo_nir}, and the corresponding model spectra are shown in Fig.~\ref{fig:nir_spectra}. See section~\ref{sec:synthesis} for a discussion regarding the abundances provided in Table~\ref{table:compo_nir}.

\begin{table*}[tbh]
\small
\begin{center}
\caption[]{\em{\small Composition of the VNIR best-fit model obtained for each asteroid\\}}
\label{table:compo_nir}
\begin{tabular}{lcccccccc}
\hline \hline \noalign {\smallskip}
Asteroid & Olivine (am.) & Pyroxene (am.) & Forsterite (cr.) & Enstatite (cr.) & Pyrrhotite & Troilite & Iron & Neutral$^{*}$ \\
\hline \noalign {\smallskip}
(24) Themis & 0.19 & 0.00 & 0.01 & 0.00 & 0.02 & 0.00 & 0.00 & 0.78 \\
(62) Erato & 0.20 & 0.00 & 0.03 & 0.00 & 0.00 & 0.00 & 0.00 & 0.77 \\
(90) Antiope & 0.41 & 0.00 & 0.19 & 0.00 & 0.00 & 0.00 & 0.09 & 0.31 \\
(171) Ophelia & 0.18 & 0.00 & 0.02 & 0.00 & 0.05 & 0.00 & 0.03 & 0.72 \\
(268) Adorea & 0.36 & 0.00 & 0.08 & 0.00 & 0.00 & 0.00 & 0.24 & 0.32 \\
(316) Goberta  & 0.12 & 0.00 & 0.00 & 0.00 & 0.07 & 0.00 & 0.00 & 0.81 \\
(379) Huenna & 0.17 & 0.00 & 0.01 & 0.00 & 0.04 & 0.00 & 0.00 & 0.78 \\
(383) Janina & 0.18 & 0.06 & 0.05 & 0.00 & 0.02 & 0.00 & 0.00 & 0.69 \\
(526) Jena & 0.07 & 0.00 & 0.00 & 0.00 & 0.00 & 0.00 & 0.00 & 0.93 \\
(621) Werdanti & 0.11 & 0.00 & 0.00 & 0.00 & 0.00 & 0.00 & 0.00 & 0.89 \\
(767) Bondia & 0.51 & 0.09 & 0.10 & 0.04 & 0.00 & 0.00 & 0.00 & 0.26 \\
(954) Li & 0.18 & 0.00 & 0.03 & 0.00 & 0.00 & 0.00 & 0.28 & 0.51 \\
\hline \noalign {\smallskip}
\end{tabular}
\begin{flushleft}
{\bf Note (1).} ${ }^{*}$Assuming optical albedo p$_v$=0.045.\\
{\bf Note (2).} The abundances provided here must be taken with caution. See discussion in section~\ref{sec:nir}. \\
\end{flushleft}
\end{center} 
\end{table*}

\subsection{Synthesis of our compositional analysis over the 0.4-2.5 and 8-13~$\mu$m ranges}
\label{sec:synthesis}

In section~\ref{sec:mir}, we found that amorphous silicates are the main contributors to the silicate emission in the MIR (8-13~$\mu$m) spectral range, suggesting that they are major components of the surfaces of the Themis family members. This is confirmed by our compositional analysis in the VNIR range (0.4-2.5~$\mu$m), where Fe-bearing olivine glasses were found to be indispensable for reproducing the spectral shape (upward concavity and local minimum at $\sim$1.3\,$\mu$m) of most family members. 
Compared to CP IDPs, where silicate glasses comprise over 50\% of the volume \citep{Bradley:1994ck, Ishii:2008jw}, we find for most members somewhat lower fractional abundances of $\sim$10\% to 30\% for the olivine glass in our models, which possibly indicates that additional fractions of spectrally neutral glasses may also be present at the surfaces of these objects.

Spectrally neutral end members were necessary components in the VNIR range to scale the spectral contrast of the olivine glass absorption band with respect to the asteroid ones. In particular, we consider that neutral carbonaceous compounds, which are widespread throughout CP IDPs, are likely present but probably do not account for all the neutral materials present on the asteroid surfaces, since this would imply abundances in several Themis members that are significantly higher ($\sim$10\% to 90\%, generally $>$60\%) than their typical abundances in CP IDPs (from $\sim$4\% to 45\%, with an average of $\sim$13\%; \citealt{Keller:1994uw}). We also found that adding small fractions ($<6\%$, except for 268 and 954 for which we find higher abundances) of pyrrhotite, troilite, and iron metal in our models helped to improve the spectral fits in the VNIR range, because they played an important role in reddening the reflectance, as well as in reducing the spectral contrast of the absorption bands. 

Crystalline silicates were found in section~\ref{sec:mir} to contribute only moderately ($<$30\%) to the asteroids' silicate emissivity in the MIR. In the VNIR, no diagnostic features around 1\,$\mu$m and 2\,$\mu$m could be identified for those silicates, although we found that adding minor fractions of these species into the models improved the fits in several cases. Such low fractional abundances agree with their typical abundances in CP IDPs where they usually account for 1\% to 5\% of the volume \citep{Bradley:1983kq}.

It should be stressed that the abundances provided here and in Table~\ref{table:compo_nir} must be taken with caution, because they correspond to a best-fit solution obtained for a specific range of particle sizes (0.1-2.0\,$\mu$m). In the optical regime considered here (i.e., where the particle size resembles the wavelength of the light), the reflectance of a given specie is strongly determined by its size. In particular, as the average size becomes smaller, the scattering efficiency decreases at longer wavelengths and provokes a spectral bluing of the reflectance (see, e.g., \citealt{Brown:2014ka}). Therefore, a redder spectrum (e.g., 268 Adorea) does not necessarily indicate a distinct mineralogy but might be caused by a different average size of the grains across the surface of the asteroid. Besides this, most species used in our models (e.g., enstatite) are fairly neutral over the VNIR spectral range at small grain sizes. As such, their presence at the surface of the asteroids cannot be ruled out by our models in this wavelength range. Overall, the modelling solutions presented here are certainly not unique, but they highlight that most Themis' spectra can be consistently reproduced by assuming a CP IDP-like composition and grain sizes. 

\begin{figure}[tbh]
\centering
\includegraphics[angle=0, width=\linewidth, trim=0cm 1cm 0cm 2cm, clip]{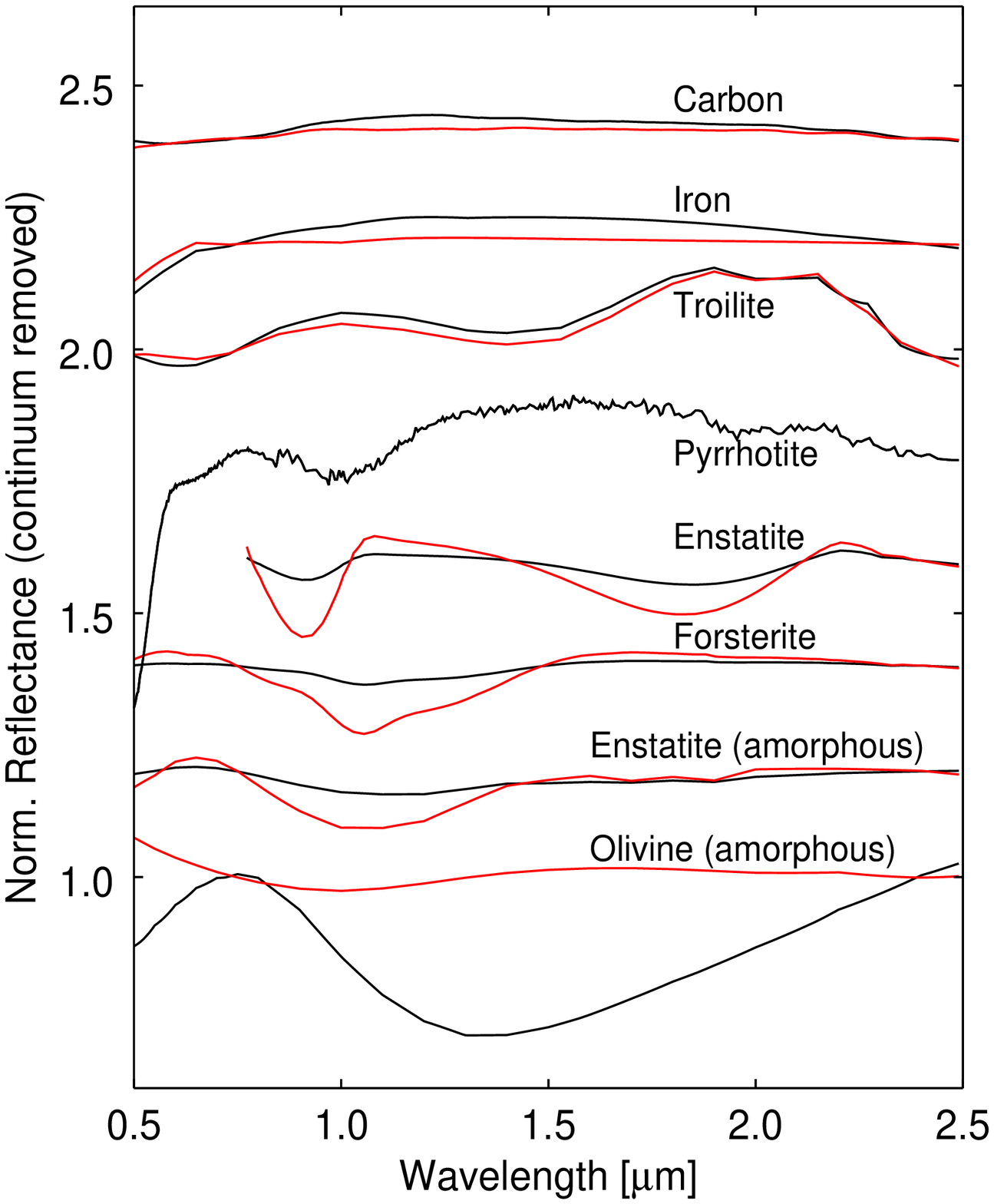}
 \caption{Reflectance spectra of the minerals used in the modelling procedure as calculated from the Mie-Hapke radiative transfer code. Spectra are normalised to unity at 0.75\,$\mu$m and offset by 0.2 unity with respect to one another for visibility. To highlight the spectral features of each specie and their relative intensities, we removed the continuum by dividing each spectrum by a line that fits the spectrum at 0.75\,$\mu$m and 2.40\,$\mu$m. The red curves correspond to a distribution of grains with large sizes compared to the wavelength (20-30~$\mu$m), whereas the black curves correspond to a distribution of grains smaller than or similar to the wavelength (0.1-2.0~$\mu$m). Since no optical constants were available for pyrrhotite, we used a reflectance spectrum from the USGS database for this mineral. See section~\ref{sec:grainsizes} for a discussion of the reflectance spectrum of amorphous olivine.}
\label{fig:minerals}
\end{figure}

\section{Discussion}
\label{sec:discussion}

In this section, we discuss the main implications resulting from these new observations and from our compositional modelling work. 

\subsection{VNIR spectral diversity}
\label{sec:sptvar}
Over the VNIR spectral range, our dataset reveals a wide spectral diversity amongst the Themis family members. This behaviour reflects either compositional heterogeneities in the original parent body or differences in the physical properties of the asteroids' regolith. A variation in the physical properties of the regolith may be due to different particle sizes at the objects' surfaces -- larger objects being more inclined to gravitationally retain finer dust -- or caused by a change in the level of hydration due to a progressive decrease in the asteroid's distance to the sun. However, our sample reveals a correlation neither between the spectral slopes and diameters ($r=0.17$) nor between the spectral slopes and the perihelion distances ($r<0.1$) (Table~\ref{table:phys}). We therefore consider the spectral variations amongst the family members to be mainly due to differences in terms of mineral composition. Finally, a few objects  (223, 461, and 468) appear to be spectrally distinct from the rest of the family, which may be attributed to fragments sampling a different compositional layer of the family progenitor (e.g., \citealt{Campins:2012km}) or to interlopers in the family. Along these lines, \citealt{Durda:2007im} predicted that 461 is an interloper on the basis of numerical simulations of the collisional event that led to the formation of the family.

\subsection{Small ($<2\,\mu$m) silicate grains as main surface components of the Themis family members}
\label{sec:grainsizes}

The identification by \citet{Vernazza:2015ei} of CP IDPs as possible analogues of the surfaces of icy asteroids (BCG-, P-, and D-types) has placed an important constraint on the typical grain size present on these objects' surfaces. At the same time, it revealed that geometric-optics calculations, which are widely used for modelling the spectral properties of solar system bodies' surfaces (e.g., asteroids, TNOs, Mars, etc.), are to a large extent invalid in the case of icy asteroids' surfaces. 

The present work has allowed highlighting the impact of small particles ($<$2~$\mu$m) on the overall reflectance in the VNIR spectral range and, in particular, the ability of small grains to produce spectral features in this wavelength range. Indeed, while Fe-bearing olivine glass is optically neutral at these wavelengths for "large" ($>$10~$\mu$m) grain sizes, we find the spectral shape for smaller grain sizes ($<$2~$\mu$m) to be consistent with the one observed for most Themis family members, that is, a broad concave-up spectral shape and a local minimum reflectance around 1.3~$\mu$m (Fig.~\ref{fig:minerals}). Such spectral behaviour is somewhat unexpected because mineral features tend to diminish with particle size in the geometric-optics regime. At particle sizes close to the wavelength and below, however, we find that the spectral behaviour changes rapidly with particle size. Such a prediction is supported by other radiative transfer models that use a statistic Monte-Carlo approach (e.g., \citealt{Pilorget:2013gb}) and that find comparable spectral behaviours for small (sub-micrometre- sized) grains (Carter et al., private communication) with the one predicted by our Mie-Hapke analytic model.

Owing to the present lack of laboratory measurements for ultra-fine grained materials ($<$2~$\mu$m), interpreting the surface composition of icy asteroids relies on the use of radiative transfer models. The acquisition in the near future of reflectance and emittance spectra for ultra-fine grained materials would be extremely valuable for refining the present predictions (both \citealt{Vernazza:2015ei}'s findings and the present work).

\subsection{CP IDP-like mineralogy for most Themis family members}
\label{sec:idp-like}

Considering grain sizes in the 0.1-2\,$\mu$m range, we find that the surfaces of most Themis family members are predominantly composed of amorphous olivine and neutral compounds (likely amorphous carbon) and of small fractions of crystalline silicates. This composition is compatible with the one measured for CP IDPs in the laboratory. The remaining major CP IDPs' building blocks (pyrrhotite, troilite, iron metal) could not be directly detected in this work, although we find that adding minor fractions of these species into the model usually improves the fit by providing a reddening of the modelled mixture.

Finally, we found three notable exceptions in our sample: (223)~Rosa, (461)~Saskia, and (468)~Lina. No good-matching models could be found for these objects, suggesting a rather distinct mineralogy from that of CP IDPs. These objects could either be interlopers or simply sampling a distinct compositional layer of the original parent body (see further discussion in section~\ref{sec:trend}).

\subsection{Implications for the thermal evolution of the Themis parent body}
\label{sec:trend}

Assuming different initial conditions for the formation of the Themis parent body, \citet{CastilloRogez:2010kya} used a geophysical model to constrain its early thermal evolution due to internal heating by the decay of short-lived radioactive nuclides (e.g., ${ }^{26}$Al). Three scenarios, hereafter (a), (b), and (c), were investigated assuming either a different initial composition or a different time of formation for the Themis parent body. Models (a) and (b) assume a formation from a homogeneous mixture of ice and rock, whereas model (c) assumes an accretion from hydrated silicates. Model (a) assumes a time of formation of 3 My after the formation of calcium-aluminium-rich inclusions (CAIs), whereas models (b) and (c) assume a time of formation of 5 My after CAIs. 

\begin{figure}[tbh]
\centering
\includegraphics[angle=0, width=\linewidth, trim=0.2cm 0cm 0.08cm 0cm, clip]{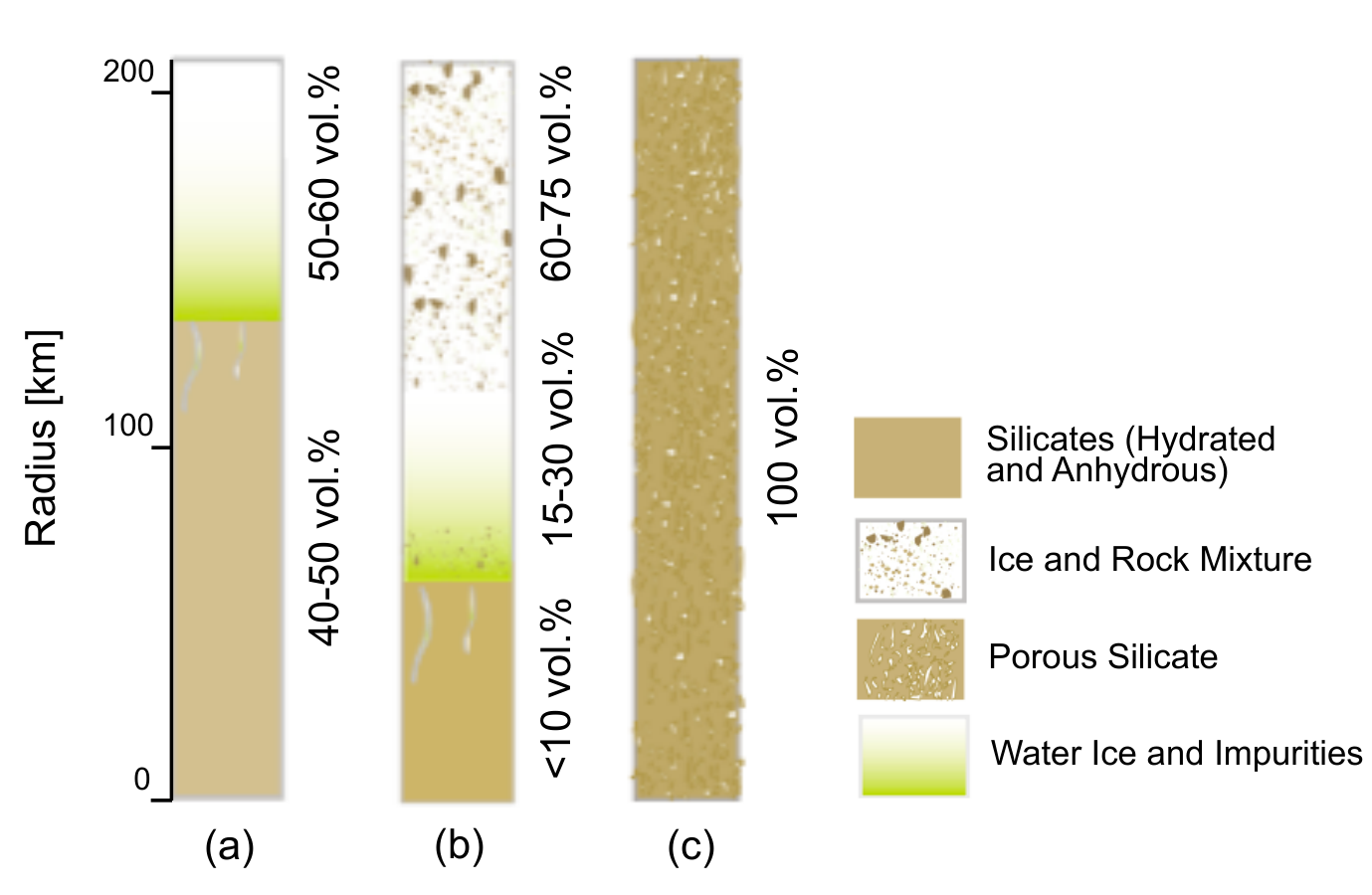}
\caption{Distribution of rock and ice throughout the Themis parent body at the time of break-up ($\sim$2\,Gyrs after its formation), assuming different initial compositions and times of formation: (a) a homogeneous mixture of ice and silicates (1:1 ratio) and a time of formation of 3~My after CAIs, (b) a homogeneous mixture of ice and silicates (1:1 ratio) and a time of formation of 5~My after CAIs, and (c) hydrated silicates and a time of formation of 5~My after CAIs. Adapted from \citet{CastilloRogez:2010kya}.}
\label{fig:castillo}
\end{figure}

Scenario (c) appears very unlikely because 1/ it can hardly explain the spectral diversity observed among the family members (the temperature reached in the innermost layers of the parent body is too low for thermal metamorphism of the silicates); 2/ in this scenario the family formation would lead to a myriad of fragments consisting of hydrated silicates, which is ruled out by our observations, and 3/ it would require all the volatiles in the family to have an exogenic origin.
The resulting compositional structure for scenarios (a) and (b) is a somewhat differentiated body with a silicate (hydrated and anhydrous) core representing between 10\% and 50\% of the body's volume and an outer shell consisting of a mixture of anhydrous silicates and ice (Fig.~\ref{fig:castillo}). The relative volumes of the core and the shell depend on the initial quantity of short-lived radioactive nuclides accumulated in the interior of the Themis parent body, which itself depends on when this body was formed. Following these two scenarios, the catastrophic collision undergone by the parent body would lead to the formation of a small number of the Themis family members that consist of a mixture of ice and anhydrous silicates, while other fragments could consist of pieces of the core, namely of hydrated silicates. 

At first glance, these two scenarios appear compatible with ours and other observations of the Themis family members:

- Previous spectroscopic studies of the Themis family members have shown that hydration is present in different forms on the objects' surfaces as oxidised iron in phyllosilicates \citep{Florczak:1999et, Kaluna:2016jo}, hydroxyl-bearing minerals \citep{Takir:2012cza}, and water ice \citep{Campins:2010fi, Rivkin:2010ge, Hargrove:2015ii}. 

- The density values recorded so far for two family members (90 Antiope and 379 Huenna: <1.3 g/cm$^3$ ; \citealt{Descamps:2007im, Marchis:2008eq}) likely imply a large amount of ice(s) in the interior of these bodies. 

- Our compositional investigation provides evidence of anhydrous IDP-like silicates in the family. That most members' surfaces appear consistent with a composition made of anhydrous silicates suggests that only the innermost layers of the parent body underwent aqueous alteration. Therefore, scenario (b) (a formation time of 5 Myrs after CAIs) appears more consistent overall with our observations than scenario (a) (a time formation of 3 Myrs after the formation of CAIs) in which aqueous alteration also occurs in the outer shell.

The observed anti-correlation between the VNIR spectral slopes of the Themis family members and their albedos (Fig.~\ref{fig:slope}) may provide additional support for a heterogeneous composition (i.e., layered internal structure) of the parent body, with the bluer and brighter members consisting of anhydrous (CP IDP-like) silicates partially covered by ices and the redder and darker members consisting of aqueously altered silicates (e.g., hydrated IDPs). A hydrous composition for the reddest members is supported in the MIR by a comparison between the spectrum of (468)~Lina and that of a hydrated IDP (Fig.~\ref{fig:nir_spectra}). Future observations in the three-micron region will help in determining whether the correlation is due to a gradient in the degree of aqueous alteration.

\section{Summary}
\label{sec:summary}
In this work, we have constrained the surface composition of a sample of Themis family members using a hybrid Mie-Hapke scattering model. 
It appears that over a wide wavelength range (0.4-13~$\mu$m), most of the family members' spectra can be consistently reproduced assuming a CP IDP-like composition and grain sizes ($<$2~$\mu$m). One object (468~Lina) appears spectrally consistent with hydrated IDPs.

The composition inferred in this work for the Themis family members suggests that the parent body accreted from a mixture of ice and anhydrous silicates (mainly amorphous) and subsequently underwent limited heating via the decay of short-lived radionuclides. By comparison with existing thermal models that assume a 400-km diameter parent body \citep{CastilloRogez:2010kya}, the accretion process of the Themis parent body must have occurred relatively late ($>$4 Myrs after CAIs) so that only moderate internal heating occurred in its deep interior, allowing aqueous alteration of the core but preventing aqueous alteration of the outer shell.

Future observations in the three-micron region will help to improve our understanding of the layering within the primordial parent body and, in particular, help quantify the extent of the aqueously altered layer, as well as the extent of the icy layer.


\begin{acknowledgement}
The authors are grateful to Humberto Campins for reviewing this paper. Observations reported here were obtained at the Infrared Telescope Facility, which is operated by the University of Hawaii under contract NNH14CK55B with the National Aeronautics and Space Administration. FED acknowledges support from the NASA's Planetary Astronomy programme under grant number NNX12AL26G.
\end{acknowledgement}


\bibliography{references}

\begin{appendix}
\section{Radiative transfer code}
\label{sec:app}

{\it Visible and near-infrared spectral range:} To model the reflected light of the asteroids in the 0.4-2.5\,$\mu$m spectral region, we used a radiative transfer code based on the Mie and Hapke light scattering theories. As end members, we used all the major and most minor minerals found in CP IDPs: crystalline olivine and pyroxene, silicate glasses, sulfides, FeNi metal, and amorphous carbon. For each material, we assumed that the number of grains per bin size obeys a power-law distribution (${\rm d}n(\alpha)=\alpha^p{\rm d}\alpha, p<0$) covering a characteristic range of sizes for IDP building blocks (we arbitrarily chose 0.1-2.0~$\mu$m). 

For each bin size, reflectance was calculated from the complex refractive index of the material ($m=n+ik$) using Hapke's equation \citep{Hapke:1993tt}: 
\begin{equation}
   r_\lambda(\mu_0, \mu, g) = \frac{w_\lambda}{4\pi} \frac{\mu_0}{\mu_0+\mu} [P(g) + H(\mu_0, w_\lambda) H(\mu, w_\lambda) -1]
\end{equation}
where $\mu_0$ and $\mu$ are the respective cosines of the incident ($i$) and emergence ($e$) angle of the light, $g$ is the phase angle between $i$ and $e$, $P(g)$ is the particle phase function, $H$ is the Chandrasekhar function, and $w_\lambda$ is the single-scattering albedo of the surface particles. In our models, the Chandrasekhar function was calculated using the two-stream approximation and an isotropic phase function was used. To limit the total number of free parameters, all the calculations were made at zero phase angle ($g=0$).

For large particles compared to the wavelength (arbitrary defined as $d>4.5\lambda$), we calculated $w_{\lambda}$ using the Hapke-derived (geometric optics, \citealt{Hapke:1993tt}) formulation: 
\begin{equation}
   w_{\lambda} = S_e+\frac{(1-S_e) (1-S_i) \Theta}{(1-S_i) \Theta}
\end{equation}
where $\Theta\,=\,e^{-\alpha\,<d>}$ is the internal-transmission factor of the particle, $\alpha\,=\,4\,\pi\,k\,\lambda^{-1}$ is the absorption coefficient of the particle, $<d>\,\approx\,0.9\,d$ is the mean distance traveled by a ray of light in the particle, and $S_e$ and $S_i$ are the coefficients for external and internal diffusion on the surface of the particle:
\begin{equation}
   S_e = \frac{(n-1)^2+k^2}{(n+1)^2+k^2}+0.05
,\end{equation}
\begin{equation}
   S_i = 1-\frac{4}{n(n+1)^2}
.\end{equation}

For small particles compared to the wavelength ($d<1.5\lambda$), we calculated $w_{\lambda}$ using the Mie theory \citep{Bohren:1983wi} for homogeneous spheres. In that case, $w_{\lambda}$ was computed as the ratio between the scattering ($\sigma_S$) and extinction ($\sigma_E$) cross sections of the particles:
\begin{equation}
   w_{\lambda} = \frac{\sigma_S}{\sigma_E}
.\end{equation}
Here, $\sigma_S$ and $\sigma_E$ were derived as series of the scattering coefficients $a_n$ and $b_n$:
\begin{equation}
\label{eq:e1}
   \sigma_S = \frac{2 \pi}{k^2}\sum_{n=1}^{\infty} (2n+1)(|a_n|^2+|b_n|^2)
,\end{equation}
\begin{equation}
\label{eq:e2}
   \sigma_E = \frac{2 \pi}{k^2}\sum_{n=1}^{\infty} (2n+1){\rm Re}\{a_n+b_n\}
,\end{equation}
which themselves were expressed in terms of {\it Riccati-Bessel} functions. 

When expanding equations \ref{eq:e1} and \ref{eq:e2} to terms of order $X^4$ (where $X = \pi d \lambda^{-1}$), one retrieves the Rayleigh approximation for the scattering and extinction cross sections:
\begin{equation}
\label{eq:e3}
   \sigma_S = \frac{2}{3} \pi d^2 X^4 \left|\frac{m^2-1}{m^2+2}\right|^2
\end{equation}
\begin{equation}
\begin{split}
\label{eq:e4}
   \sigma_E = \pi d^2 X\,{\rm Im}\Bigg(\frac{m^2-1}{m^2+2}\bigg(1+\frac{X^2}{15}\,\frac{m^2-1}{m^2+2}\,\frac{m^4+27m^2+38}{2m^3+3}\bigg)\Bigg)\\+\frac{2}{3}\pi d^2 X^4 \,{\rm Re}\Bigg(\bigg(\frac{m^2-1}{m^2+2}\bigg)^2\Bigg)
\end{split}
.\end{equation}

The Rayleigh approximation is valid under the condition that $X<<1$ (i.e., for particles that are very small compared to the wavelength). When $X\approx1$, equations \ref{eq:e1} and \ref{eq:e2} must be expanded to higher terms of X. 

Finally, we defined a transition spectral region ($1.5\lambda<d<4.5\lambda$) where $w_\lambda$ was calculated as a weighted average of its values as derived from the two theories:
\begin{equation}
   w_{\lambda} = x\,w_{\lambda_{Mie}} + (1-x)\,w_{\lambda_{geo-optics}}
\end{equation}
so that $x=1$ at $d=1.5\lambda$, $x=0$ at $d=4.5\lambda,$ and x varies linearly in the intermediate spectral region. \\

{\it Mid-infrared spectral range:} The mid-infrared emissivity of crystalline silicates and minerals strongly depends on the geometry of the particles. To compute the emissivity at these wavelengths, we therefore chose to assume a distribution of randomly oriented hollow spheres \citep{Min:2003km} with small sizes compared to the wavelength (X$\ll$1), and internal vacuum inclusions ranging from 0\% to 90\%. 

The emissivity of each asteroid was fitted by a linear combination of the absorption efficiency $Q_A$ of the end members, that is, the ratio between the absorption cross section $\sigma_A$, where $\sigma_A = \sigma_E - \sigma_S$, and the "physical" cross section $\pi (d/2)^2$ of the particles.

In the MIR, $\sigma_E$ and $\sigma_S$ were derived from equations \ref{eq:e3} and \ref{eq:e4}, except for a small modification to account for the geometry of the grains. In equations \ref{eq:e3} and \ref{eq:e4}, the term $(m^2-1)(m^2+2)^{-1}$ -- sometimes referred as the "polarisability per unit material volume" ($p$) -- is only valid for homogeneous spheres. In the case of hollow spheres, this term was replaced by

\begin{equation}
        p=\frac{(m^2-1)(2m^2+1)}{(m^2+2)(2m^2+1)-2(m^2-1)^2f}
\end{equation}

\noindent where $f$ is the internal volume fraction of void ranging from 0\% to 90\% in our models.

\end{appendix}

\end{document}